# Potential Global Sequestration of Atmospheric Carbon Dioxide by Drylands Forestation

Murray Moinester[1*], Joel Kronfeld[2]

[1]School of Physics and Astronomy, Tel Aviv University, 69978 Tel Aviv, Israel

[2]Department of Geophysics, Tel Aviv University, 69978 Tel Aviv, Israel

E-mails:

murray.moinester@gmail.com; https://orcid.org/0000-0001-8764-5618

joel.kronfeld@gmail.com

*Correspondence: Murray Moinester

Drylands forestation offers the potential for significant long-term sequestration of atmospheric $CO_2$. Here we consider sequestration of both organic and inorganic carbon by a planted semi-arid forest, based on carbon that originates from atmospheric $CO_2$. Measurements at Israel's Yatir forest give a sequestration rate of ~550 g $CO_2$ $m^{-2}$ $yr^{-1}$ as organic carbon in the trees' biomass. In addition, based on Yatir measurements, an inorganic carbon precipitation rate is estimated to give an additional 216 g $CO_2$ $m^{-2}$ $yr^{-1}$ globally, via calcite ($CaCO_3$) precipitation in soil. This soil inorganic carbon (SIC) sequestration is due to a combination of microbial activity on organic soil carbon, and the formation of soil carbonic acid ($H_2CO_3$) that arises from the reaction of soil water with $CO_2$ exhaled from tree roots. Significantly, low rainfall in drylands precludes dissolving precipitated calcite. Published estimates restrict the potential drylands surface available for sustainable forestation to ~4.5 million $km^2$, only ~10% of the global drylands. The dominant limitation is the apparent lack of water. However, immediately under many drylands, there are paleowaters (fossil water) that had recharged underlying aquifers during prior wetter climatic regimes. Conservatively, including fossil water, at least ~9.0 million $km^2$ is available for afforestation. Measurements at Yatir show that drip irrigation to 18% average Soil Moisture (higher than the rainfed 12% SM) would double the organic carbon sequestration rate. In addition, the tree density could be increased, which would independently double the organic carbon sequestration rate. The potential total annual sequestration rate is then conservatively estimated as 20.0 Gt $CO_2$ $yr^{-1}$, divided between 14.0 Gt $CO_2$ $yr^{-1}$ (organic) and 6.0 Gt $CO_2$ $yr^{-1}$ (inorganic). This corresponds to 100% of the annual rate of atmospheric $CO_2$ increase

## 1 Introduction

The twin ecological problems of increasing global warming and ocean acidification are inextricably intertwined with increasing levels of atmospheric carbon dioxide. $CO_2$ is currently being emitted globally at roughly 40 billion tons per year (40 Gt $CO_2$ $yr^{-1}$). About 50% of these emissions accumulate in the atmosphere, 25% in the ocean, and 25% on land (Friedlingstein 2020, NOAA 2025). Since the Industrial Revolution, the $CO_2$ concentration in the atmosphere has risen from ~280 ppmv to ~427 ppmv at present (ProOxygen 2025). The global atmospheric $CO_2$ reservoir of ~3200 billion tons is presently increasing annually by ~20 billion tons. This increase is occurring mainly through the burning of fossil fuels, forest fires and deforestation. The released $CO_2$ is a potential cause of both increased global warming and ocean acidification (Pilson 1998). Indeed, the global Conference of the Parties climate pacts call for the world's nations to significantly reduce $CO_2$ emissions, to prevent (by 2030) global warming from rising more than 2.0 °C above pre-industrial levels. Slowing the rise in global temperatures also requires simultaneously proactively reducing atmospheric $CO_2$ concentrations. Relatedly, very expensive, large climate engineering projects have been proposed by $CO_2$ removal (CDR) and solar radiation management (SRM) (Linnér and Wibeck 2015).

To date, $CO_2$ greenhouse gas and its potential for global warming get most of the media attention. However, the problem of increasing ocean acidity has a direct bearing on oceanic health and the global food supply. The pH of the ocean has been decreasing since the pre-industrial period due to increasing levels of atmospheric $CO_2$ (Pilson 1998). The ocean has so far served as a sink for roughly 25% of $CO_2$ released to the atmosphere. $CO_2$ when combined with $H_2O$ forms $H_2CO_3$, carbonic acid. The increasing oceanic carbonic acid concentration has led to increased ocean acidity. This in turn inhibits the ability of the oceans to absorb more atmospheric $CO_2$. The pH of ocean water is presently 8.1 (NOAA 2025), which represents a 30% increase in acidity since the pre-industrial era. This increase correlates with anthropogenic releases of $CO_2$ (Joos 2023). The increased acidity puts stress on planktonic organisms that build their shells from $CaCO_3$ (calcium carbonate or calcite), as well as on other marine life forms (NOAA 2025). Solving this problem requires reducing the atmospheric $CO_2$ concentration in contact with the ocean surface.

Thus, the solution to both ecological problems requires the removal of atmospheric $CO_2$, coupled with its long-term storage. Such a solution may be found within the carbon cycle, whereby soils and trees abstract atmospheric $CO_2$ and store large stocks of global carbon. Simply, trees take in $CO_2$ through the process of photosynthesis, which converts the carbon to organic carbon which makes up almost half of a tree's mass. Organic matter is transferred to the soil by tree litter, and decomposition of fallen trees, aided by microbial action to form soil organic matter (SOC). Atmospherically derived $CO_2$ can also be removed by forests transforming this gas into inorganic carbon, either as soil inorganic carbon (SIC) or as dissolved inorganic carbon (DIC). We elaborate on the processes below. Forestation is a simpler and less expensive method for removing atmospheric carbon than massive and expensive high tech engineering projects. But where should this needed large-scale planting be carried out? Boysen et al. (2017) and Ostberg et al. (2018) pointed out that attaining mandated climate goals via temperate zone forestation would overtly reduce arable lands available and needed for food production. Moreover, large amounts of fertilizer would be required, whose runoff could degrade water supplies.

Drylands may be more suitable than temperate lands for afforestation efforts to remove atmospheric $CO_2$ by organic and by inorganic carbon sequestration. Drylands make up over 40% of the world's surface (Reynolds 2007), covering almost 45 million $km^2$. This includes semi-arid, arid, hyper-arid sub-humid regions. Drylands regions are generally less populated and provide low agricultural and economic value. They would be prime regions for forestation were it not that the climate is hot and harsh, and that sufficient water appears to be lacking. Thus, most of these regions have not been previously considered for forestation (Minnemeyer 2014; Rohatyn 2022). For example, the latter reference assumed that only 10% of the drylands area could sustain forests.

Moreover, in climate mitigation modeling, there is a playoff between two first-order climate-influencing factors: the reflectivity of the land and the amount of $CO_2$ in the atmosphere. The higher the reflectivity (albedo), the less is the heating due to the incoming solar radiation. The arid lands are regions of high albedo. By planting trees, the albedo decreases as dark vegetation replaces reflective landscapes. The shift in reflectively results in greater radiative heating which may override the cooling effect of atmospheric $CO_2$ reduction; despite higher evapotranspiration, or increased cloud cover (Luo 2024).

Thermal emission in the infrared wavelength range from the Earth back to space is crucial for the Earth's energy balance and resultant equilibrium average temperature. $CO_2$ in the atmosphere efficiently intercepts and captures some of this infrared energy, and then re-emits part of that energy back to the Earth. Thereby, heat is trapped on the Earth, which increases the Earth's equilibrium temperature. This is popularly known as the greenhouse effect. That is, $CO_2$ greenhouse gas contributes to warming as its concentration rises; alternatively, a decline in its levels would lead to a cooling effect. Note however that without the approximately 34°C temperature increase due to greenhouse gases, the Earth would be very cold instead of an average observed global temperature of ~15°C. That is, the human contribution is only responsible for changing the greenhouse gas warming from 34°C to say 36°C. It is this extra 2°C that is popularly associated with global warming.

We start with a preview of our own estimate of global drylands carbon sequestration. Rohatyn et al. (2022) estimated that 4.5 million $km^2$ of semi-arid and dry sub-humid drylands are potentially forestable. Qubaja et al. (2020) measured an organic sequestration rate in Yatir forest (with no irrigation or fertilization) of 150 grams of carbon per $m^2$ per year, which equates to 550 grams of $CO_2$ per $m^2$ annually. This translates to 2.5 billion tons of organic $CO_2$ per year over an area of 4.5 million $km^2$. Adding an estimated 1.0 billion tons of inorganic $CO_2$ per year (described in Sections 5 and 6) gives 3.5 billion tons $CO_2$ sequestered per year. Taking into account the presence of fossil water (described in Section 2), we may double the potential usable area to 9.0 million $km^2$. Assuming the same organic and inorganic sequestration rates as in the non-irrigated, rainfed Yatir forest, and assuming that Yatir is representative, this area could then sequester approximately 7.0 billion tons $CO_2$ per year. But drip irrigation to 18% average Soil Moisture (higher than the rainfed 12% SM) would increase the global carbon sequestration rate to ~14.0 Gt $CO_2$ $yr^{-1}$ (Qubaja 2025). With irrigation, the tree density could also be doubled (Qubaja 2025). The global carbon sequestration rate may then be estimated as ~20.0 Gt $CO_2$ $yr^{-1}$, ~100% of the annual rate of atmospheric $CO_2$ increase (Qubaja 2025).

We note that removing 20.0 Gt $CO_2$ from the atmosphere would disturb the equilibrium between atmosphere, soil and oceans; approximately 50% of $CO_2$ emitted to the atmosphere presently accumulates in the atmosphere, 25% in the ocean, and 25% on land (NOAA 2025). Equilibrium restored by soils and oceans emitting $CO_2$ to the atmosphere; resultant effective $CO_2$ reductions per year would be 10.0 Gt, 5.0 Gt, and 5.0 Gt for atmosphere, oceans, soils**.** Considering this expected equilibration, a very important additional benefit of forest carbon sequestration would be a reduction in ocean acidity.

## 2 Fossil water under the harshest of deserts

It is precisely under the harshest drylands that water can be economically found for afforestation. Across North Africa, there are extensive extant water reserves in fossil groundwater. These can be found in areas where there is virtually no effective recharge today, such as under the Sahara Desert. The Nubian Sandstone Aquifer System is the largest of these, with an aerial extent exceeding 2 million $km^2$ under the Sahara, extending under Egypt, Libya, Chad and Sudan. The total amount of freshwater in the aquifer is estimated as ~370 thousand $km^3$ (CEDARE 2000), about 100,000 times Israel's Sea of Galilee (Lake Kinneret). The latest major recharge of this aquifer was ~128 thousand years ago (Godfrey-Smith 2008). The aquifer has correlative stratigraphic sections with concomitantly large water reserves under central Sinai, the Negev desert in Israel, parts of Jordan and Arabia. The potential use of fossil water aquifers for drylands irrigation is under current investigation (Kronfeld 2025). These aquifers show little or no surface expression as to their true potential to sustain long-term forestation. By exploiting them, much larger land areas with a concomitant local water supply would be available for afforestation than had been previously assumed (Rohatyn 2022; Minnemeyer 2014). Indeed, paleowater filled aquifers can be found in semi-arid and arid areas world wide (Verhargen 1991). Due to the low population densities and inhospitable climate, the hydrology and ages of the waters in many such aquifers underlying deserts globally have not been fully investigated.

## 3 Yatir forest study area and methods of study

Israel's planted Yatir forest is a 28 km$^2$ Aleppo pine forest growing at the semi-arid timberline with no irrigation or fertilization (KKL-JNF 2022). Jewish National Fund (Keren Kayemeth LeIsrael) foresters have planted ~4 million trees at Yatir since 1964. This site (GPS: 31°20′ N, 35°03′ E) is situated above the carbonate Mountain Aquifer at an elevation of ~650 m above sea level, at the edge of Israel's Negev desert. It is the largest forest in Israel (KKL-JNF 2022).

The mean annual precipitation is ~285 mm, falling solely during the winter rainy season as high intensity rain events, while the remainder of the year is hot and dry. The mean annual potential evaporation and precipitation are 1,600 mm and 285 mm, respectively. The runoff is negligible. The groundwater table is located at greater than 300 m depth (Qubaja 2020A, 2021). Conditions are at the drier and hotter limit (Aridity Index AI = 0.18) compared to the world's semi-arid (AI = 0.2-0.5) region (Arora 2002). Despite the present low precipitation, the Yatir forest is productive and stores organic carbon relatively effectively (Grünzweig 2003).

## 4 Organic sequestration

Organic carbon sequestration is based on utilizing plant photosynthesis to abstract atmospheric $CO_2$ and then storing it as organic carbon in trees. Photosynthesis during daylight drives biological pumps, whereby atmospheric $CO_2$ entering leaf stoma combines with $H_2O$ to produce organic carbon, glucose ($C_6H_{12}O_6$), as well as $O_2$ oxygen as a byproduct. This process can be represented by the following equation:

$$6CO_2 + 6H_2O \rightarrow C_6H_{12}O_6 + 6O_2 \quad (1)$$

Trees utilize this glucose to form biomass, such as wood and leaves. The reverse reaction, either by direct oxidation or involving catabolic reactions, releases the stored carbon either directly to the atmosphere or into the soil profile by root exhalation:

$$C_6H_{12}O_6 + 6O_2 \rightarrow 6CO_2 + 6H_2O \quad (2)$$

The organic carbon residence time for this sequestration (Eq. 1) is at least 150 years, considering the typical 100-150 year life span of Aleppo pines and their long decomposition time after falling. Thus, extensive afforestation has been proposed as being effective in sequestering atmospheric $CO_2$, both as Aboveground Biomass Carbon (ABC), as well as in the roots (Johnson and Coburn 2010; Kell 2012; Tans and Wallance 1999; Watson 2000). The carbon is in the form of wood above ground (trunk, branches, bark) and below ground (roots), leaf litter, and tree products (nuts, acorns, fruit). Approximately 50% of a tree's biomass is composed of carbon.

Israel's Yatir Forest was found to have an organic carbon sequestration (OCS) rate of 150 grams per square meter per year, corresponding to 550 g organic $CO_2$ $m^{-2}$ $yr^{-1}$. This extrapolates to a global 5.0 billion tons $CO_2$ per year in 9.0 million km$^2$. The value 550 g organic $CO_2$ $m^{-2}$ $yr^{-1}$ sequestration rate (OCS) was obtained using the data of a 15 year long monitoring program that combined eddy covariance (EC) flux measurements (Weizmann Institute Group), as well as carbon stock counting inventories. The EC method is a key atmospheric measurement technique employed to determine net vertical forest-atmosphere exchange fluxes of $CO_2$, water vapor, heat, etc. Uncertainties (~20%) were assessed by comparing stock-based and flux-based approaches (Qubaja 2020B). Ecosystem-scale accounts of carbon stocks (CS) were estimated in permanent study plots involving tree size parameters and soil organic carbon (SOC) and then converted to CS with allometric equations and soil sample analyses (Parresol 1999; West 1997). This stock-based approach is generally performed on a few small plots that need to be scaled up. The original forest inventory assessment was performed in 2001 (Grünzweig 2007) and was repeated 15 years later (Qubaja 2020B) on the same five 30 × 30 $m^2$ plots in the central part of the forest. A detailed recent forest inventory included estimates of the four main components: standing biomass, litter, soil, and removal (mortality, thinning, and sanitation). The mean annual net ecosystem productivity based on CS over the 15-year observation period represents the average increase in carbon in soil and tree CS over the observation period. We assume in the discussion that the organic carbon sequestration (OCS) rates at Yatir forest are representative of global drylands.

A drip irrigated plot (24% SM) in Yatir showed approximately 3 times higher organic $CO_2$ sequestration than a comparable rainfed plot (Qubaja 2025). We assume irrigation to 18% SM, lower than that of 24% SM in the irrigated plot, to reduce dissolving precipitated calcite. With irrigation, the tree density could also be doubledfrom 1 tree per 6X6 $m^2$ to to 1 tree per 4X4 $m^2$ (Qubaja 2025). Assuming an irrigation enhancement of approximately 3 would increase the global organic carbon sequestration rate from ~5.0 Gt $CO_2$ $yr^{-1}$ to ~14.0 Gt $CO_2$ $yr^{-1}$ (Qubaja 2025).

**5 Inorganic sequestration associated with CO2 gas exhalation**

The Yatir forest's inorganic carbon sequestration data includes gas and soil samples collected in depth profiles extending from the surface to a maximal depth of ~4.5 m. Carbon isotope ratios ($C^{13}/C^{12}$ and $C^{14}/C^{12}$) were measured as a function of depth in the liquid and solid phases of soil profiles in the unsaturated zone (USZ), and then presented in standard $\delta^{13}C$ and $\Delta^{14}C$ notation (Carmi 2015). The old residual limestones have a characteristic marine signature of $\delta^{13}C$ = ~0 ‰. The $\delta^{13}C$ of the bicarbonate ($HCO_3^-$), originating from isotopic depleted root-exhaled $CO_2$ ($\delta^{13}C$ ~ −26 ‰, C3 plants, Clark and Fritz 1997) is slowly enriched as a function of depth by exchange with the relict marine carbonate to approximately half this value in the USZ. Radiocarbon is a unique tracer for labeling atmospheric derived sources of carbon gas, liquid and solid carbon phase within the USZ. The host sediment initially contains no $^{14}C$. This radioactive carbon isotope is produced in the upper atmosphere. Allogenic calcite, besides its $\delta^{13}C$ signature, is readily distinguished from pedogenic carbonate soil calcite in that it contains no radiocarbon ($\Delta^{14}C$ = -1000, zero percent modern carbon). By tracking carbon isotopes ($^{12}C$, $\delta^{13}C$, $\Delta^{14}C$) as a function of depth in the liquid and solid phases of soil profiles in the USZ, it was demonstrated that the source of a significant portion of the precipitated calcite was originally $CO_2$ respired from tree roots (Carmi 2019). A tree's roots exhale $CO_2$ into the soil after some of the tree's glucose (produced by photosynthesis) has been oxidized to supply energy for the tree's cellular processes.

The $CO_2$ in the soil gas of the USZ can attain partial pressures many times above the ambient atmospheric $CO_2$ partial pressure (Clark and Fritz, 1997). This facilitates the reaction:

$$CO_2\ (gas) + H_2O \rightarrow H_2CO_3 \rightarrow H^+ + HCO_3^-, \qquad (3)$$

in which soil $CO_2$ combines with soil moisture to form a carbonic acid solution, which rapidly dissociates to $H^+$ and bicarbonate ($HCO_3^-$). When the dissolved inorganic carbon (DIC) concentration increases (due in part to evaporation), the concentration exceeds the solubility of calcite. The DIC (mainly bicarbonate) and the soluble bivalent cations (mainly $Ca^{2+}$) then combine to form and precipitate pedogenic calcite within the USZ:

$$Ca^{+2} + 2HCO_3^- \rightarrow CaCO_3 \downarrow + CO_2 \uparrow + H_2O. \qquad (4)$$

Rate limiting factors for these reactions include the abundance of $Ca^{2+}$ (or $Mg^{2+}$), and the partial pressure of $CO_2$. Considering Eq. 4, it had been previously accepted (Monger 2015) that if the calcium ions were not derived from silicates, no atmospheric $CO_2$ would be sequestered. This would follow if for every mole of calcite formed, one mole of $CO_2$ is returned to the atmosphere. While Eq. 4 would suggest such a conclusion, it does not describe the actual situation in the soil column. In Israeli soils, as in soil of many regions, there are exogenous sources of bivalent cations, particularly $Ca^{2+}$, that include calcium ions desorbed from the exchange sites on clays imported by air-borne dusts (Singer 2007). In addition, calcium can be brought in by rain or sea spray; and its concentration in the soil would depend on the distance from the coast (Loewengart 1964). Moreover, the Eq. 4 reaction takes place within the USZ soil column, which is generally thick in semi-arid regions. Only the topmost part of the USZ is in direct contact with the atmosphere. Gas from this location most likely does diffuse out of the soil. However, $CO_2$ released lower in the soil column is more likely to enter and mix with the relatively high partial pressure $CO_2$ in the USZ. The high pressure facilitates its reaction with soil moisture and the formation of bicarbonate, which then combines with calcium or magnesium ions (Carmi 2019). Where rainfall is plentiful, as in temperate zones, this precipitate dissolves. In drylands regions, where rainfall is sparse, precipitated calcite can remain stable for millennia (Cerling 1984).

Certain soil microbes, including specific bacteria and fungi, engage in metabolic processes that significantly influence mineral precipitation, particularly the formation of calcite (CaCO3). These microorganisms utilize various organic compounds as energy sources. During the metabolism of organic matter, they exhale carbon dioxide (CO2) as a byproduct, which can accumulate in the soil environment.

The techniques developed and used for sampling the soil moisture and the inorganic carbon via isotopic measurements are presented in Carmi et al. (2007, 2009). The rate of rainwater infiltration at Yatir was measured as ~11 ± 2.2 cm $yr^{-1}$ using Tritium in water as a tracer (Carmi 2015). Water including $T_2O$ is extracted from soil water samples of a sediment profile as a function of depth (Kaufman 2003). The depth profiles were converted to time profiles, using the measured infiltration rate. Combined with the data from the mid-core depth (2.2 meters) as representative, the calcite deposition rate into the sediment was found to be 22 gram atmospheric $CO_2$ per year per cubic meter of sediment (Carmi 2019), interpreted as the inorganic carbon sequestration (ICS) rate. We assume a 6 meter global average depth of root respiration in drylands USZ; even though tree roots extend downwards to much greater depths in drylands compared to temperate zones (Canadell 1996). Microbial respiration of $CO_2$ contributes to the measured rate. The estimated global sequestration rate in the estimated 9.0 million $km^2$ is then $6\times9.0\times10^{12}\times22 = 1.2 \times10^{15}$ gram per year or 132 g $m^{-2}$ $yr^{-1}$ or ~1.2 Pg $yr^{-1}$, ~1.2 billion metric tons per year (Qubaja 2025). The precipitated calcite is very stable in low rainfall drylands regions. The drip irrigated plot (24% SM) in Yatir showed approximately 3 times higher organic CO2 sequestration than the rainfed plot (Qubaja 2025). We assume irrigation to 18% SM, lower than that of 24% SM in the irrigated plot, to reduce dissolving precipitated calcite. With irrigation, the tree density could also be doubled. Assuming therefore an approximately 3 times irrigation enhancement yields the drylands inorganic sequestration rate associated with $CO_2$ exhalation to be ~3.5 Pg $CO_2$ $yr^{-1}$ (Qubaja 2025).

**6 Direct Microbial sequestration of atmospheric $CO_2$ as inorganic calcite**

Soil microbes have long been recognized for their ability to precipitate calcite (Boquet, 1973). These microorganisms utilize diverse and intricate biochemical pathways to convert atmospheric $CO_2$ into calcite, a processes that has been extensively documented (Zhu and Dittrich, 2016; Jaing, 2022). McCutcheon et al. (2014) noted that microbial abstraction of atmospheric $CO_2$ can be a long-term and cost-effective strategy. More recently, Liu et al. (2020) demonstrated that desert microbes can directly convert atmospheric $CO_2$ into insoluble calcite precipitates under optimal laboratory conditions. Their research revealed a calcite precipitation rate of 52 μg C per kg of soil per day, translating to approximately 70 g $CO_2$ $m^{-3}$ $yr^{-1}$, assuming a soil density of 1.2 g $cm^{-3}$. This rate is about three times higher than the previously reported inorganic sequestration rate of 22 g $CO_2$ $m^{-3}$ $yr^{-1}$. Field studies by Zheng et al. (2022) confirmed the presence of calcite-forming microbes in dryland farm soils.

Several factors inhibit microbial activity and calcite precipitation in drylands, including intense UV irradiation affecting the top 10 cm of soil, limited moisture, and reduced organic carbon content (Chen, 2023). Forestation in drylands, coupled with improvements in soil nutrition and moisture, may significantly enhance deeper microbial activity (Cao, 2018; Guo, 2018). Soil microbes, including bacteria and fungi, play a vital role in decomposing organic matter such as dead trees and leaves. This decomposition process releases $CO_2$ into the soil, where it dissolves in soil water to form bicarbonate ($HCO_3^-$), which subsequently combines with calcium ions ($Ca^{2+}$) to precipitate as calcite. The calcium ions are sourced from the weathering of rocks, soil minerals, and the release of calcium from clays.

In mature forest ecosystems, direct microbial activity contributes to the calcite precipitation rate. Scheibe et al. (2023) illustrated that forestation contributes nutrients and carbon to the soil, stimulating microbial activity. However, representative soil profiles in dryland forests reflecting the concentration of calcite-forming microbes are still lacking. Assuming a soil depth of 1.2 meters and approximately 70 g $CO_2$ $m^{-3}$ $yr^{-1}$ across 9 million $km^2$, an additional 0.8 billion tons of $CO_2$ could potentially be sequestered annually as calcite. Furthermore, with an irrigation enhancement of about threefold, the inorganic sequestration rate associated with direct microbial calcite production in drylands could increase to approximately 2.5 Pg $CO_2$ $yr^{-1}$ (Qubaja, 2025). Over time, microbial communities may migrate deeper through the soil profile, thereby proportionally increasing their contributions to calcite production.

Extracellular polymeric substances (EPS) are crucial to the direct calcite precipitation process. Composed of complex mixtures of polysaccharides, proteins, lipids, and nucleic acids produced by microorganisms, EPS plays several roles in mineral precipitation. Polysaccharides are a primary component, providing structural integrity and enhancing mineral binding. Functional proteins within EPS engage in chemical reactions and act as enzymes that facilitate these mineral interactions. Lipids and nucleic acids contribute to the stability and overall functionality of EPS.

Many bacteria and fungi synthesize EPS in response to environmental stressors or as part of their natural growth cycles. The types and amounts of EPS produced can vary based on microbial species and environmental conditions, such as nutrient availability and pH levels. EPS is vital for mineral precipitation, particularly in the initial formation of calcite, as it provides surfaces for calcium carbonate crystals to nucleate. Negatively charged sites on EPS attract positively charged calcium ions ($Ca^{2+}$), which increase the local concentration of these ions. EPS also helps stabilize minerals and modify local pH conditions, promoting calcite precipitation. Once calcite crystals begin to form, EPS can trap and stabilize them, preventing their dissolution.

The accumulation of calcite facilitated by microbial EPS enhances soil structure by binding soil particles together, thereby improving aeration and water retention. This process contributes to the long-term storage of carbon in soils. Thus, EPS plays a vital role in how microorganisms interact with their environment, supporting their survival and essential geological and ecological processes such as calcite formation. Through the production of EPS, microbes significantly influence soil health, nutrient dynamics, and carbon cycling.

Thus, through the interplay of EPS microbial metabolism, the availability of calcium ions, and appropriate environmental conditions, soil microbes effectively facilitate the direct formation and precipitation of calcite in nutrient-rich soils. This process plays a significant role in enhancing soil health, promoting carbon cycling, and stabilizing soil structure.

The contributions of the EPS process versus $CO_2$ exhalation by roots and microbes in dryland forests to calcite precipitation can vary significantly based on specific environmental conditions and the types of microbial communities present. It is estimated that the EPS process contributes approximately 30% of the total calcite precipitation in dryland soils, while the remaining 70% can be attributed to $CO_2$ produced by microbial respiration. This $CO_2$ combines with soil water and bivalent calcium ions, resulting in calcium carbonate formation through increased bicarbonate concentration, leading to subsequent calcite crystallization. The specific percentages of contributions can vary based on local conditions, microbial communities, and environmental factors. More accurate estimates require field studies to assess these contributions in specific locations.

The depth at which extracellular polymeric substances (EPS) formation occurs in soils varies significantly between ecosystems, such as drylands and temperate forests. Soil type plays a crucial role; in sandy soils commonly found in drylands, EPS may be less effective at greater depths due to lower water retention and nutrient availability. Moisture availability is vital for microbial activity and EPS production. In drylands, limited water restricts microbial growth and confines EPS formation to the upper soil layers. Temperature also affects microbial metabolism and activity. Extreme temperatures in drylands may limit microbial activity. Organic matter content significantly impacts EPS formation. High organic matter supports robust microbial populations. In drylands, lower organic matter reduces microbial activity and EPS production. Additionally, the chemical composition of the soil, including pH, influences microbial activity. Certain microorganisms prefer specific pH ranges, impacting the depth of EPS production. Generally, neutral to slightly acidic soils favor microbial growth, allowing the EPS process to extend deeper.

Disturbances such as tilling, erosion, or the activity of soil fauna can disrupt microbial communities and affect EPS formation. Disturbances in drylands can limit this depth. Typically, EPS activity occurs within the top 10-30 cm of soil due to moisture limitations and lower organic matter content. Overall, the effective depth of the EPS process is driven by factors such as soil type, moisture availability, climate, organic matter content, pH, and disturbances. Generally, EPS activity is more pronounced and extends deeper in moist, organic-rich temperate forests compared to arid drylands.

The percentages of calcite precipitation due to the extracellular polymeric substances (EPS) process versus the CO2 exhalation by microbes in dryland forests can vary significantly based on specific environmental conditions and the type of microbial communities present. It is often estimated that the EPS process contributes approximately 30% of the total calcite precipitation in dryland soils. The remaining 70% of calcite precipitation can be attributed to CO2 produced by microbial respiration, which combines with soil water and bivalent calcium ions. This process results in the formation of calcium carbonate through increased bicarbonate concentration and subsequently calcite crystallization. The specific percentages can vary based on local conditions, microbial communities, and environmental factors. More accurate estimates require field studies to assess the contributions in specific locations.

**7 Total Global Inorganic sequestration**

We therefore estimate the ICS rate in global drylands as ~6.0 Pg CO2 $yr^{-1}$ assuming root and microbial exhalation up to depths of 6 meters, and direct soil microbial production of calcite to 1.2 meter. These two ICS contributions were described above as ~3.5 Pg $CO_2$ $yr^{-1}$ and ~2.5 Pg $CO_2$ $yr^{-1}$ respectively. The global OCS rate was estimated above as ~14.0 Gt $CO_2$ $yr^{-1}$. The total global carbon sequestration rate is then estimated as ~20.0 Gt $CO_2$ $yr^{-1}$, ~100% of the annual rate of atmospheric $CO_2$ increase (Qubaja 2025).

**8 Contaminants in surface and groundwater that would restrict their use only to afforestation**

Although the waters in several fossil aquifers have contaminants that make them inappropriate for domestic or animal husbandry purposes, they are nonetheless well suited for planted trees. For example, waters in the phreatic aquifers of the western Kalahari are marked by high nitrate values, well above maximal permissible human health levels, as observed in 18% of the measured wells (Heaton 1984). However, no adverse effects would accrue to trees irrigated with this water. A more egregious example is the soluble radium contamination, found above maximal permissible levels for human consumption, within the water of the huge Nubian Sandstone Aquifer System as well as its extensions across the northern Sinai and into the Negev desert (Vengosh 2009; Sherif 2021). Elevated levels of this pernicious radioelement have also been encountered locally in the fossil waters of the extensive Saq aquifer, the principal water supply of Saudi Arabia (Faraj 2009). Yet, these waters could be used in wide-scale afforestation efforts (Tripler 2014). The reason is that desert plants have developed biochemical barriers to non-nutrients, resulting in small soil to plant Transfer Coefficients (Zafrir 1992). Arsenic, a known carcinogen, is another contaminant that should preclude the use of arsenic contaminated water for domestic purposes. The large Indus River Valley in Pakistan, which covers an area of 176,000 km², would be a harsh desert if not for the Indus River that flows through it. The river gains its fresh water from the melt of glaciers in the mountains at its source. This densely populated, economically vital area is dependent upon the river waters for irrigation, for its fish industry, and for growing diverse agricultural crops. Unfortunately, the waters and the soils of the young underlying phreatic aquifer are naturally polluted with arsenic (Podgorski 2017). The arsenic is transferred to its agricultural produce as well as bio-accumulated in the fish (Shah 2009). Arsenic incorporated into the diet of the inhabitants (Arain 2009) is slowly poisoning them. Trees would not suffer from these arsenic levels, as would humans and their livestock. This otherwise arid area may best use the extensive available but contaminated water for afforestation projects.

**9 Sustainability of forestation**

To what extent can afforested trees successfully grow in harsh arid environments, even if water is supplied? A very high mortality was noted for example, ranging from 20-80%, depending upon the species, for reforested seedlings in Anatolia, Turkey (Yildiz 2022). Cao et al. (2010) described how to avoid high mortality rates. They suggest using trees that are known to be adapted to the local environment. There are many species that would be suitable, particularly as underlying fossil waters would obviate the need to be dependent solely on rainfall. Many can thrive on marginal soils, while at the same time improving soil properties, particularly with increases in soil organic matter, soil carbon and nitrogen. Some of the following trees can be successfully planted in arid zones. The deep-rooted species of the Tamarix and the Acacia families (both have roots that can extend to depths in excess of 30 m). These two and the desert willow (*Chilopsis linearis*) are hardy desert plants. The mesquite trees (*Prosopsis*) have even deeper tap roots (up to 50 m). The deep rooting zone would facilitate deep $CO_2$ penetration for weathering and inorganic carbon sequestration as discussed above. Afforestation can stabilize soil and prevent erosion, offer food for wildlife, and enrich soil with nitrogen. It can also provide high value crops. These would include the Physic Nut (*Jatroph curcas*), which has extensive medicinal uses, the Jojoba (*Simmondsia chinensis*) for its pharmacological and industrial oils; and the Moringa (*Moringa oleifera*) used as a nutrient supplement and for its antioxidant and anti-inflammatory properties. The palm (*Phoenix dactyliferia*), when irrigated, have been profitably planted in dense groves in hyper-arid regions of Israel to provide high quality dates for export (Abobatta 2019; Cadman 2024; Canadell 1996; Wiser 2024). Thus, if groundwater is available, the arid and hyper-arid regions should not be dismissed for their forestation potential. In this regard, previous afforestation models that increased the area of total drylands available for afforestation to 15% (Potapov 2011) or to 25% (Bastin 2011) should merit reconsideration. We assume conservatively that a drylands area of 9 million km$^2$ (20%) is feasible for forestation.

One proposed reforestation method would require manpower efforts by local farmers, but not large financial investments. This method is known as **F**MNR, Farmer Managed Natural Regeneration (FMNR 2025, Chesire 2025, Walker 2024). FMNR is a sustainable land management practice developed by environmental conservationist Tony Renaudo that empowers local farmers to restore and manage trees by nurturing existing root systems and promoting natural regeneration. It fosters the regrowth of native trees that have been cut down or damaged. Chesire et al. (2025) has reviewed worldwide FMNR activities. The below ground root system, which comprises a significant fraction of the organic carbon, has been previously been assumed to die off, to decompose, to begin returning its stored carbon to the soil and atmosphere. However, the roots do not necessarily die off quickly. Even in harsh, desert-like conditions, the root systems of many felled trees remain alive, comprising "underground forests". Tree stumps, supported by their live roots, often spontaneously grow new sprouts. Usually, without human intervention, only a small bush develops. However, by carefully pruning and protecting new sprouts, selecting the tallest and strongest ones, a new tree can grow quickly. FMNR is applicable in the world’s drylands, where it can enhance soil fertility, improve moisture retention, reduce erosion, and promote the regeneration of native trees. It has already transformed millions of hectares of degraded land in at least 29 countries. A significant portion of the world's drylands is certainly suitable for forest regeneration. Following forest regeneration, carbon sequestration is renewed.

FMNR has demonstrated that drylands are suitable for both forestation and agriculture. Moreover, it has been found that the mixing of forests with agriculture is mutually beneficial. The same forested area can grow both trees and agricultural food products. The trees provide shade protection, preserve soil moisture, provide soil stability, increase biodiversity, provide lumber and improve agricultural productivity. The agricultural activity in turn provides fertilizer to the trees.

**10 Dryland carbon sequestration cooling and albedo reduction reassessed**

The sustainable area for potential global sequestration of atmospheric $CO_2$ by drylands forestation was conservatively doubled to 9.0 million $km^2$ with respect to the area considered suitable by Rohatyn et al. (2022). Based on this larger area, we make a new assessment as to how much $CO_2$ can be removed from the atmosphere and stored by dryland afforestation. We obtain a potential total sequestration rate of ~20.0 billion tons $CO_2$ $yr^{-1}$.

This rate would correspond to a significant 100% of the annual increase of 20 billion tons of $CO_2$ presently being added to the global atmospheric $CO_2$ reservoir of ~3200 billion tons. It can be directly applied to attenuating the rate of ocean acidification. However, the desired global cooling effect by atmospheric $CO_2$ reduction would be partially offset by the expected reduction in a forest's land surface albedo. The reason is that the albedo (reflectivity) of the incident solar radiation on dark forests would be reduced (leading to a warmer surface) compared to the pre-forestation, higher albedo, cooler desert surface. The effect of the change in land surface albedo has been known for some time to affect land surface temperatures, which in turn could significantly affect precipitation both locally and over extended distances (Otterman 1974; Charney 1975). Many investigations have been carried out recently on the effects of forest albedo as it pertains to global warming (Luo 2024; Rothore 2022; Yosef 2018; Zhang 2022; Zhang Li 2022; Wang 2024). The impact of albedo is nuanced. Miralles et al. (2025) contend that forests are not only a product of their environment, but also modulate climate itself. For example, forests release biogenic, volatile organic compounds which are important in forming nuclides of raindrops, cloud formation, and the enhancement of cloud reflectivity. This ultimately can lead to the formation of low-lying dense clouds over the forest, which would inhibit the penetration of short-wave (UV and visible) solar radiation. Forest evapotranspiration cools while supplying humidity to the environment. The trees add roughness to the landscapes, which results in a turbulent air flow aiding in the dissipation of surface energy and daytime cooling (relative to non-forested ground). Williams et al. (2021) noted that the loss of forested land can lead to changes in temperature values that range from a net warming to a net cooling. Wang et al. (2024) suggest that previous estimates of the albedo effect may be overestimated by a factor of ~3.

Healey et al. (2025) have studied albedo variables such as tree species and forest age in a wide scale study of forests spanning across the entire USA. On average, they found that albedo has offset almost 50% of the cooling benefit attributable to organic carbon sequestration. They found more specifically that the land surface albedo of a forest is strongly affected by both tree species and age. Such albedo changes are due to differences in leaf/foliage color and thickness properties, canopy structure, growth patterns, and surface reflectivity; specific to how different tree species change as they mature. For example, for Lodgepole pine forests, they found a decreased carbon storage benefit of ~40% in young (10 years) forests; while the sequestration benefit diminishes completely after 100 years. This suggests that tees should be harvested at ages 10-20 years. Besides, global warming needs to be addressed over a 10 year time scale before considering what happens after 100 years. More studies are certainly needed of the albedo effect, to help choose the optimal trees for different global forests. Throughout, we use an estimated albedo effect sequestration reduction of ~30% (Rohatyn 2022). This value warrants further study and may be revised accordingly.

For a 20.0 Gt $yr^{-1}$ total rate, due to the decreased albedo, the total $CO_2$ "equivalent" atmospheric cooling sequestration rate would be reduced to 14.0 Gt $yr^{-1}$ (Rohatyn 2022). This value would still correspond to a significant 70% of the annual increase of 20 billion tons of $CO_2$ presently accumulating in the global atmosphere. Again, this is a preliminary and rough estimate. But it justifies extending the afforestation potential into the arid and hyper-arid lands. Liang et al. (2024) have advanced the idea that forestation affects ecology in a dynamic way. Irrigating and using the land will influence the local ecology by increasing soil moisture. The trees would add organic matter to the soil, carbon and nitrogen, which in turn beneficially affects the microbial populations. The evaporation of water not only has a local cooling effect (King 2024); but combined with evapotranspiration, the moisture may stimulate cloud formation. The resulting dense low-lying clouds, which are strongly coupled to the ground, would reflect solar radiation. This would increase the effective albedo and thus contribute to cooling of the land, notwithstanding lost reflectivity. The added moisture in turn may eventually spur increased precipitation. Estimating more precise effective sequestration rates would require considering the choice of tree species, stand densities, canopy cover, precipitation levels, cloud cover; and how $CO_2$ concentrations equilibrate between atmosphere, forested drylands and oceans. Forested lands may, over time, increase their carbon sequestration potential while extending the forests' natural ranges.

## 11 **Limitations**

Our global sequestration estimate is based on measurements at a single location, on which basis we extrapolate to the entire planet. The validity of the OCS and ICS extrapolations depends therefore on how representative Yatir forest is of the global drylands. For ICS, it also depends on the validity of the estimated average global root depth (6 m), and the assumed microbial activity depth (1.2 m). With an aridity index AI = 0.18, drier and hotter than most of the world's semi-arid regions (AI = 0.2-0.5), Yatir is nonetheless healthy and productive without irrigation or fertilization. Therefore, its sequestration rate is possibly lower than the average drylands rate. We cautiously but boldly estimate that our global extrapolation serves as a minimum threshold, with the possibility of the rate being higher.

Certainly, taking, including, and averaging data from other drylands forests in varied geographical locations would be more representative. One should include forests covering the full range of drylands aridity indices, as well as having a variety of topographic conditions such as slope or soil type or soil thickness. Since this was beyond the scope of the current study, our estimated global sequestration rate is very tentative. Our objective is not to give a definitive result, but rather to generate interest such that other research groups may be motivated to provide complementary data and to thereby refine our initial rough estimate.

The extrapolation we present is based on a planted Aleppo Pine (*Pinus halepensis*) forest. After such a forest reaches equilibrium, maybe 100 years after planting, the quantity of organic carbon fixed by photosynthesis will be approximately equal to the quantity released. The release takes place via decay of fallen trees and litter, natural and human initiated fires, harvesting of forest products, and respiration of $CO_2$ as trees convert stored carbohydrates into energy. Overall, these processes represent the dynamic balance of carbon cycling within a mature forest. We emphasize here that the positive sequestration rate of organic carbon from tree growth is valid only during the first 50-100 years of the forest's life until it reaches equilibrium. The actual time for planted forests in general would depend on the tree type and specific management practices and environmental conditions. By contrast, over the entire lifetime of the forest, including after equilibration, there is a continuous increase in the inorganic carbon sequestration that is permanently fixed in the form of calcite.

## **12 Discussion**

Can forest management be designed so that the organic sequestration rate would continue long term? During the initial years, rapid tree growth can lead to significant carbon sequestration. But we noted above that a forest typically reaches a state of equilibrium where organic carbon inputs and outputs are roughly equal, after ~50 years. However, implementing a sustainable harvesting strategy, where a fraction of trees is harvested annually and replaced with new plantings, should promote ongoing organic carbon sequestration. This practice, often referred to as selective logging or continuous cover forestry, would help maintain a long term balance between carbon uptake and release. Planting a diverse array of tree species is also important to enhance resilience and productivity. Different species can have varying growth rates, root systems, and ecological roles, which can contribute to greater overall carbon storage. In this regard, fire resistant trees should get priority to reduce forest fire damage. In addition to above-ground biomass, improving soil health through practices like mulching, reduced tillage, and organic amendments can enhance soil carbon storage. Regular monitoring of growth rates, carbon stocks, and forest health will allow adapting management practices based on changing conditions in order to ensure the continued effectiveness of the forest's carbon sequestration capability. By combining the above strategies, a sustainable forest management plan can be developed that encourages long-term organic carbon sequestration while supporting biodiversity and ecosystem health.

Such sustainable forest management practices that focus on long-term organic carbon sequestration have been implemented in various places around the world. Two notable examples are Agroforestry and Sustainable Forest Management (SFM). Agroforestry systems that combine agriculture and forestry can enhance carbon sequestration while providing economic benefits by growing crops alongside trees. Countries like Canada and Sweden have adopted SFM practices that include selective logging and maintaining or enhancing forest biodiversity, which helps to ensure ongoing carbon storage while allowing for timber production.

**13. Carbon Credits**

Irrigated drylands forest planted with diverse types of trees could be eligible for carbon credits. To qualify for credits, the forest must demonstrate measurable carbon sequestration. This involves calculating the amount of $CO_2$ captured by the trees over time. There are various carbon credit certification standards, such as the Verified Carbon Standard (VCS), the Gold Standard, and the Climate Action Reserve. They all have specific requirements for forest projects, including baseline scenarios (what would have happened without the project) and monitoring procedures. The project would need to adhere to the guidelines set by the chosen carbon credit program, including having a well-defined project plan, legal title to the land, and metrics for measuring carbon stocks and changes over time. Regular monitoring and reporting are essential components of carbon credit programs. It would be necessary to track tree growth and any changes in carbon stocks, which may require third-party verification. Carbon credits are typically issued for projects that ensure a long-term commitment to maintaining the forest and its carbon stocks for a designated period, usually 20 years or more. Projects that promote biodiversity and provide additional co-benefits (such as improved water quality or enhanced local livelihoods) would be more attractive to carbon credit programs and buyers. The global drylands forestry project we propose should qualify for carbon credits.

**14 Conclusions**

Global drylands offer extensive area for forestation efforts aimed at reducing global warming and oceanic acidity by removing atmospheric $CO_2$ through the combination of inorganic and organic carbon sequestration. Our estimate of the potential quantity of sequestered $CO_2$ offers optimism for the efficacy of the project, yet should be considered both conservative and tentative. Several parameters related to inorganic sequestration for example are not well constrained at present. These involve microbial calcite precipitation rates, and the sequestration of atmospheric $CO_2$ in the form of bicarbonate in underlying aquifers.

We nevertheless estimate the potential for significant long-term removal of atmospheric $CO_2$ via carbon sequestration in global drylands forests. Our estimate is based on measurements in the planted semi-arid Yatir forest in Israel. This forest sequesters organic carbon in the trees' biomass and inorganic carbon precipitated as stable calcite ($CaCO_3$) in the forests' soils.

The potential maximal efficacy of global forestation for reducing global warming and ocean acidification depends on the maximal area available for sustainable forestation. The dominant limitation in global drylands regions is the apparent lack of sufficient water. Published evaluations ostensibly restrict the potential drylands surface available for sustainable forestation to ~4.5 million $km^2$, only ~10% of the global drylands. However, immediately underlying many drylands, plentiful groundwater is available. These are paleowaters (fossil waters) that have recharged underlying aquifers during prior wetter climatic regimes. Conservatively, taking these fossil waters into account, we estimate that an area of at least ~9.0 million $km^2$ would be available for drylands afforestation.

We then estimated the ICS rate in global drylands as ~6.0 Pg CO2 $yr^{-1}$ assuming root and microbial exhalation up to depths of 6 meters, and direct soil microbial production of calcite to 1.2 meter. These two ICS contributions were described above as ~3.5 Pg $CO_2$ $yr^{-1}$ and ~2.5 Pg $CO_2$ $yr^{-1}$ respectively. The global OCS rate was estimated above as ~14.0 Gt $CO_2$ $yr^{-1}$. The total global carbon sequestration rate is then estimated as ~20.0 Gt $CO_2$ $yr^{-1}$, ~100% of the annual rate of atmospheric $CO_2$ increase.

However, the cooling effect through removing the $CO_2$ greenhouse gas is more nuanced, as these forests would alter surface energy dynamics. The transformation of bright high albedo (reflectivity) drylands to darker forests could reduce the positive projected climate cooling significantly for the mass of greenhouse gas removed from the atmosphere. Considering only the reduced land surface albedo, the effective cooling sequestration rate may be closer to ~14.0 Gt $CO_2$ $yr^{-1}$. However, some positive atmospheric feedback mechanisms may oppose this reduced albedo cooling. For example, increased forest evapotranspiration may decrease surface temperature and increase cloud formation. The species and age of the trees as well as the canopy height may mitigate the heating effects of reduced albedo. And thick, low lying stratocumulus clouds would tend to decrease heating by incident solar short-wave radiation.

All things considered, drylands forestation comprises a more cost-effective, long-term use of drylands and fossil water when considering cooling benefits; relative to the current profligate, agricultural exploitive draining of these fossil aquifers. A signal case in point is the hydrologic situation in the Ogallala or High Plains paleo-aquifer, covering 450 thousand $km^2$ of the central semi-arid USA (McMahon 2004; Nativ and Smith 1985). Present use combined with past uncontrolled exploitation and waste has seriously depleted this aquifer (Little 2009; Scanlon 2012). Similar examples of excessive use of fossil water has been taking place in Saudi Arabia (De Nicola 2015), Arizona (Stanley-Becker 2023), Yemen (Aljawzi 2022) and Ethiopia (Pearce 2012). Therefore, we suggest that these ongoing agricultural practices should be curtailed; and less water-intensive ways of utilizing the irrigated water should be implemented rather than water intensive agricultural growing of fruits or vegetables. Forestation should have priority in all drylands areas where paleo-aquifers are the dominant water supply.

To date, drylands regions are generally not industrialized, and are too dependent on rainfall to offer more than employment in only marginally profitable agriculture or herding. Persistent poverty may be a focus for political or social instability. The sequestration described here is restricted to drylands regions. Far from being a drawback, this gives added economic potential to these marginal regions. Besides, in addition to sequestering carbon, semi-arid forests provide important ancillary environmental functions. These include preventing encroaching desertification, producing oxygen, increasing precipitation (Yosef 2018), reducing ocean acidification, improving soil structure and quality and soil stability, reducing erosion and runoff, reducing soil biogenic nitric oxide emissions, reducing air particulate pollution, providing lumber and charcoal, providing wildlife habitat and recreational facilities, providing forest management employment to local populations, and producing carbon offset credits to be sold on global carbon trade exchanges. Considering all these advantages, one may expect widespread social acceptance by the local population; this being an important requirement for the success of the project.

FMNR, Farmer Managed Natural Regeneration, may foster the regrowth of native trees in deforested drylands. The root systems of many felled trees remain alive, comprising "underground forests". Tree stumps often spontaneously grow new sprouts. By carefully pruning and protecting new sprouts, new trees can grow quickly in a significant portion of the world's drylands. Following forest regeneration, carbon sequestration would be renewed. FMNR may also possibly be applied in areas where trees have been burned in forest fires, both in drylands and temperate regions. Fires that spread quickly by heavy winds may, for example, be more heat damaged around the canopy compared to near ground level. In such cases, cutting selected burned trees at about 0.5 meter above the ground may promote the regeneration of sprouts from the remaining trunk, particularly if the root system is intact. This approach would be especially effective for tree species that are resilient to fire. Additionally, deforested temperate areas may also be appropriate for FMNR, especially if the soil remains fertile and remnants of tree roots are still present to support regeneration. Expanding on these subjects is beyond the scope of the present study.

With the availability of fossil water and the judicious choice of trees adapted to aridity, the drylands regions of the planet should be the prime focus for global forestation efforts. Sequestration of atmospheric $CO_2$ would also reduce the rate at which oceanic acidification is occurring. Overall, our estimate demonstrates the global potential for drylands forests to develop into significant carbon sinks. Note, however, that a significant additional area of potential global forest dryland—on the order of millions of square kilometers—may be available for FMNR reforestation (FMNR 2025, Cheshire 2025, Walker 2024), beyond the 9 million square kilometers used in the $CO_2$ sequestration calculations above. Thus, the 9 million square kilometers included in our sequestration estimate should be considered a conservative value. We emphasize the need for further drylands forest sequestration measurements, and the need to begin implementing a global land management policy of widespread afforestation and reforestation regeneration in drylands regions.